\documentclass[
reprint,
superscriptaddress,
showkeys,
aps,
pra,
longbibliography,
nofootinbib
]{revtex4-2}

\usepackage[utf8]{inputenc}

\usepackage[hyperref]{xcolor}
\definecolor{goethe-blau}{cmyk}{1.0,0.2,0.0,0.4}
\definecolor{hellgrau}{cmyk}{0.04,0.04,0.05,0.02}
\definecolor{sandgrau}{cmyk}{0.12,0.09,0.13,0.0}
\definecolor{dunkelgrau}{cmyk}{0.25,0.25,0.30,0.75}
\definecolor{emo-rot}{cmyk}{0.04,1.0,0.8,0.07}
\definecolor{purple}{cmyk}{0.08,1.0,0.3,0.36}
\definecolor{senfgelb}{cmyk}{0.01,0.25,1.0,0.05}
\definecolor{gruen}{cmyk}{0.62,0.4,0.87,0.09}
\definecolor{magenta}{cmyk}{0.08,0.86,0.12,0.12}
\definecolor{orange}{cmyk}{0.0,0.7,1.0,0.04}
\definecolor{sonnengelb}{cmyk}{0.0,0.12,0.95,0.0}
\definecolor{helles-gruen}{cmyk}{0.4,0.17,0.81,0.07}
\definecolor{lichtblau}{cmyk}{0.8,0.0,0.06,0.04}

\usepackage[
colorlinks,
pdfpagelabels,
breaklinks,
pdfstartview=FitH,
bookmarksopen=true,
bookmarksnumbered=true,
bookmarksopenlevel=2,
plainpages=false,
hypertexnames=false,
citecolor=emo-rot,
linkcolor=goethe-blau,
urlcolor=purple,
pdftitle={},
pdfauthor={},
pdfkeywords={}
]{hyperref}

\usepackage{bm,dsfont}
\usepackage{tensor}
\usepackage{amsmath,amssymb,amsthm}
\usepackage{physics}
\usepackage{upgreek}
\usepackage[bb=boondox]{mathalfa}
\usepackage{mathtools}

\usepackage{bookmark}
\usepackage{graphicx}
\usepackage[caption=false]{subfig}
\usepackage{dcolumn}
\usepackage{slashed}
\usepackage{multirow}

\usepackage{ulem}

\usepackage{tikz}
\usetikzlibrary{decorations.pathreplacing}
\usetikzlibrary{positioning}

\def\GeV{\,\mathrm{GeV}}
\def\MeV{\,\mathrm{MeV}}

\usepackage[acronym]{glossaries-extra}
\glssetcategoryattribute{acronym}{nohyperfirst}{true}
\setabbreviationstyle[acronym]{long-short}

\newacronym{UV}{UV}{ultraviolet}
\newacronym{IR}{IR}{infrared}
\newacronym{RG}{RG}{renormalization group}
\newacronym{fRG}{fRG}{functional renormalization group}
\newacronym{LPA}{LPA}{local potential approximation}
\newacronym{SSB}{SSB}{spontaneous symmetry breaking}
\newacronym{QMM}{QMM}{quark-meson model}
\newacronym{QCD}{QCD}{quantum chromodynamics}

\allowdisplaybreaks

\newcommand\numberthis{\addtocounter{equation}{1}\tag{\theequation}}

\def\d{\,\mathrm{d}}

\newcommand{\orcid}[1]{\href{https://orcid.org/#1}{\includegraphics[height=1.9ex,width=1.9ex]{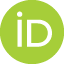}}}

\usepackage[capitalise,sort,compress]{cleveref}
\Crefname{section}{Sec.}{Sections}
\Crefname{table}{Tab.}{Tables}

\begin{document}

\title{Analysis of regulator and cutoff artifacts in the phase diagram \\ of the quark-meson model}

\author{Jonas~Stoll~\orcid{0000-0001-8204-9876}}
\email{jonas.stoll@tu-darmstadt.de}
\affiliation{
	Technische Universität Darmstadt, Department of Physics, Institut für Kernphysik, Theoriezentrum,\\
	Schlossgartenstraße 2, D-64289 Darmstadt, Germany
}

\author{Niklas~Zorbach~\orcid{0000-0002-8434-5641}}
\email{niklas.zorbach@tu-darmstadt.de}
\affiliation{
	Technische Universität Darmstadt, Department of Physics, Institut für Kernphysik, Theoriezentrum,\\
	Schlossgartenstraße 2, D-64289 Darmstadt, Germany
}
\author{Lutz~Kiefer~\orcid{0009-0002-3666-5664}}
\email{lkiefer@itp.uni-frankfurt.de}
\affiliation{Institut f\"ur Theoretische Physik, Goethe University,
Max-von-Laue-Straße 1, D-60438 Frankfurt am Main, Germany}

\author{Fabrizio~Murgana~\orcid{0009-0003-1000-9551}}
\email{fabrizio.murgana@dfa.unict.it}
\affiliation{Department of Physics and Astronomy, University of Catania, Via S. Sofia 64, I-95125 Catania, Italy}
\affiliation{INFN-Sezione di Catania,
Via S. Sofia 64, I-95123 Catania, Italy}
\affiliation{Institut f\"ur Theoretische Physik, Goethe University,
Max-von-Laue-Straße 1, D-60438 Frankfurt am Main, Germany}

\affiliation{Helmholtz Research Academy Hesse for FAIR, Campus Darmstadt, D-64289 Darmstadt, Germany}

\author{Jens~Braun~\orcid{0000-0003-4655-9072}}
\email{jens.braun@tu-darmstadt.de}
\affiliation{
	Technische Universität Darmstadt, Department of Physics, Institut für Kernphysik, Theoriezentrum,\\
	Schlossgartenstraße 2, D-64289 Darmstadt, Germany
}
\affiliation{
		Helmholtz Research Academy Hesse for FAIR, Campus Darmstadt,\\
		D-64289 Darmstadt, Germany
}
\affiliation{ExtreMe Matter Institute EMMI, GSI, Planckstra{\ss}e 1, D-64291 Darmstadt, Germany}

\author{Dirk~H.~Rischke~\orcid{0000-0002-8110-3209}}
\email{drischke@itp.uni-frankfurt.de}
\affiliation{Institut f\"ur Theoretische Physik, Goethe University,
Max-von-Laue-Straße 1, D-60438 Frankfurt am Main, Germany}
\affiliation{
	Helmholtz Research Academy Hesse for FAIR, Campus Riedberg, Max-von-Laue-Straße 12, D-60438 Frankfurt am Main, Germany
}

\begin{abstract}
We study regulator and cutoff artifacts in the quark-meson model at finite temperature and quark chemical potential within the functional renormalization-group approach using the local potential approximation.
To this end, we discuss the concept of renormalization-group consistency in effective models, which necessitates a nontrivial parameter-fixing procedure to enable a meaningful comparison of results obtained with different regulators and cutoffs.
We employ a standard range of cutoff values used in phenomenological studies and regulators that differ significantly in their analytic properties as well as in their classification according to the principle of strongest singularity. 
We find that regulator and cutoff dependences are small at low temperatures and quark chemical potentials.
At high temperatures and low quark chemical potentials, significant cutoff artifacts arise, whereas the properties of the regulator affect the dynamics in the regime governed by a chiral phase transition of first order at low temperatures and high quark chemical potentials. 
\end{abstract}


\maketitle

\section{Introduction}\label{sec:intro}
The chiral phase transition of the theory of the strong interaction, \gls{QCD}, is an area of active research, see, e.g., Refs.~\cite{HotQCD:2019xnw,Borsanyi:2020fev,Rennecke:2016tkm,Braun:2020ada,Aarts:2020vyb,Kotov:2021rah,Bernhardt:2023hpr,Braun:2023qak,Ding:2023oxy,Pisarski:2024esv,Mitra:2024mke,Fejos:2024bgl,DAmbrosio:2025ldv}.
In particular at intermediate to high baryon densities, its precise nature and order are crucial for understanding the properties of, e.g., neutron stars. 
Of particular interest is the potential existence of a first-order phase transition and an associated critical endpoint. 
However, first-principles studies of this region of the phase diagram remain challenging. 
While recent progress has been achieved with functional methods, see, e.g., Refs.~\cite{Lu:2025cls,Fu:2019hdw,Gunkel:2021oya}, lattice Monte Carlo simulations still face significant limitations at nonzero densities due to the sign problem. 
Hence, effective models remain valuable instruments for gaining at least qualitative insights into the \gls{QCD} phase structure, see, e.g., Refs.~\cite{Herbst:2010rf,Carignano:2016jnw,Rennecke:2016tkm,Berges:1998sd,Schaefer:2004en,Ihssen:2023xlp,Roder:2003uz,Eser:2015pka}. 
In particular, from a \gls{fRG} perspective, effective hadronic degrees of freedom emerge dynamically from gluon-induced multi-fermion interactions in the \gls{RG} flow with properties varying with the temperature and chemical potential~\cite{Braun:2017srn,Braun:2018bik,Braun:2019aow,Fukushima:2021ctq}.
This provides a direct connection between effective models and fundamental \gls{QCD} dynamics.
Unfortunately, effective models are typically subject to regulator and \gls{UV} cutoff dependences due to their finite \gls{UV} extent, see, e.g., Ref.~\cite{Braun:2018svj} for a discussion on more general grounds.

In this work, we assess and quantify regulator and \gls{UV} cutoff dependences within the \gls{fRG} framework, focusing on the \gls{QCD}-inspired \gls{QMM} with two quark flavors and three colors, which is introduced in \cref{sec:model}. 
In \cref{sec:FRGandEFT}, we discuss the \gls{fRG} approach and the concept of \gls{RG} consistency~\cite{Braun:2018svj}, i.e., the independence of observables from the \gls{UV} cutoff and the regulator. 
In our calculations, we employ the so-called \gls{LPA}, which corresponds to the leading order of the derivative expansion and already takes into account bosonic fluctuation effects. 
This approximation together with a discussion of the choice of regulators and \gls{UV} cutoffs considered in our analysis is discussed in \cref{sec:LPA}. 
In \cref{sec:flow-eq}, we then present the flow equation for the scale-dependent effective potential. 
In our calculations with different regulators and \gls{UV} cutoffs, we focus particularly on a consistent parameter-fixing procedure which is essential for meaningful comparisons, i.e., to assess \gls{RG} consistency of our results. 
This is discussed in detail in \cref{sec:param-fix}. 
In \cref{sec:results}, we finally present our results for the phase structure in the plane spanned by the temperature and the quark chemical potential. 
Our analysis shows that, at high temperature and small quark chemical potential, the variation of physical observables with respect to a variation of the \gls{UV} cutoff within values typically employed in phenomenological studies is stronger than the dependence on the regulator. 
At low temperature and high chemical potential, the situation is reversed. 
There, the dependence on the regulator is found to be stronger than the \gls{UV}-cutoff dependence. 
Our conclusions can be found in \cref{sec:conc}.

\section{Model}\label{sec:model}
The \gls{QMM} contains quark and meson fields. 
The quark fields are denoted by $\bar{\psi}$ and $\psi$ with $d_\gamma=4$ Dirac, $N_\mathrm{f}=2$ flavor, and $N_\mathrm{c}=3$ color degrees of freedom. 
The quarks are assumed to form mesonic pairs in the scalar-pseudoscalar interaction channel. 
The meson sector comprises $N_\mathrm{m}=4$ real fields $\phi=(\sigma, \pi_1,\pi_2,\pi_3)$, with the scalar isosinglet $\sigma$ meson and the pseudoscalar isotriplet of pions $\vec{\pi}$.
For the sake of simplicity, we suppress all indices and running variables in the following, whenever they are clear from the context.
In $d=4$ Euclidean spacetime dimensions, the \gls{QMM} is defined by the action~\cite{Braun:2009si,Tripolt:2017zgc,Braun:2018svj,Otto:2022jzl,Schaefer:2006ds} 
\begin{align*}
	\label{eq:bare_action}\numberthis
	S[\Phi] &= \int_x \Big\{ \bar\psi \mathrm{i} \big[ \slashed{\partial} -\mu \gamma_0 + h (\sigma + \mathrm{i} \vec\tau \cdot \vec\pi \gamma_\mathrm{ch})\big] \psi \nonumber\\
	&\;\;\;\; +\tfrac{1}{2} \phi \cdot (-\partial^2)\phi + \frac{m^2}{2} \phi \cdot \phi + \frac{\lambda}{4} (\phi \cdot \phi)^2 - c \sigma\Big\} \, ,
\end{align*}
where $\vec\tau$ is the vector of Pauli matrices acting in flavor space and $\gamma_0$ and $\gamma_\mathrm{ch}$ are the zeroth and chiral Dirac gamma matrices, respectively, operating in Dirac space. 
The squared mass parameter $m^2$, the quartic coupling $\lambda$, the Yukawa coupling $h$ and the parameter $c$ are adjustable model parameters that have to be tuned to match a given set of observables, such as the pion decay constant,
the pion mass,
and the constituent quark mass.
Our choice for the model parameters is discussed in detail in \cref{sec:param-fix}. 

For notational convenience, we define a super field 
\begin{align*}
	\label{eq:multi-field}\numberthis
	\Phi=(\bar{\psi},\psi,\phi) \, .
\end{align*}
Temperature is introduced via compactification of the Euclidean time direction, with 
\begin{align*}
	\numberthis \label{eq:spacetime-integral}
	\int_x \equiv \int_0^\beta \d \tau \int_{\mathds{R}^{d-1}}\d^{d-1} \vec{x} \, , \quad \text{where} \quad T = \frac{1}{\beta} \, ,
\end{align*}
and imposing periodic and antiperiodic boundary conditions for the meson and quark fields, respectively.
Furthermore, we include a (quark) chemical potential $\mu$ in \cref{eq:bare_action}, which acts as a Lagrange multiplier coupled to the conserved quark bilinear $\bar{\psi} \gamma_0 \psi$ 
and allows us to tune the value of the net baryon number density. 
The net baryon number is a conserved quantity, which results from an invariance of our action under global $\mathrm{U}(1)_\mathrm{V}$ phase transformations
\begin{align*}
	\label{eq:phase-sym}\numberthis
	\psi \mapsto u \psi \,, \quad \quad \quad \bar{\psi} \mapsto \bar{\psi} u^{-1} \, ,
\end{align*}
where $u \in \mathrm{U}(1)$. The meson parts of the action are unaffected by this transformation.

In the chiral limit, i.e., for $c=0$, the action is also invariant under global chiral $\mathrm{SU}(2)_\mathrm{L} \times \mathrm{SU}(2)_\mathrm{R}$ transformations~\cite{Schaefer:2006ds,Carignano:2016jnw}, which act as
\begin{align*}
	\label{eq:chiral-sym}\numberthis
	\psi &\mapsto (a \mathcal{P}_+ + b \mathcal{P}_-) \psi \,, \quad  \bar{\psi} \mapsto \bar{\psi} (b^{-1} \mathcal{P}_+ + a^{-1} \mathcal{P}_-) \,, \\
	\phi &\mapsto A(a,b) \cdot \phi\,, 
\end{align*}
where $a,b \in \mathrm{SU}(2)$ operate in flavor space and $\mathcal{P}_\pm = \tfrac{1}{2} (\mathds{1}_{d_\gamma} \pm \gamma_\mathrm{ch})$ are the chiral projectors. 
The transformation matrix $A(a,b) \in \mathrm{SO}(4)$ of the meson fields is induced by $a,b$.
Note that a nonzero value of~$c$ explicitly breaks the chiral symmetry.

\section{Functional renormalization group and effective models}\label{sec:FRGandEFT}
In the \gls{fRG} formalism, an infrared regulator $R_k$ is introduced into the partition function.
This regulator is parametrized by the \gls{RG} scale $k$ and is constructed to regularize modes of the partition function with momenta $p^2 \lesssim k^2$, whereas modes with momenta $p^2 > k^2$ remain largely unaffected. 
This modification of the partition function leads to the Wetterich equation~\cite{Wetterich:1992yh},
\begin{align*}
	\label{eq:wetterich-equation}\numberthis
	\partial_t \Gamma_k [ \Phi ] = \frac{1}{2}\,\mathrm{STr} \Big[  \big( \Gamma^{(2)}_k [ \Phi ] + R_k \big)^{-1} \cdot \partial_t R_k \Big] \, .
\end{align*}
where $\partial_t = - k \partial_k$ is the logarithmic \gls{RG}-scale derivative, $R_k$ is the regulator matrix, and $\mathrm{STr}$ denotes the super-trace, which includes integration over spacetime, summation over internal degrees of freedom, and a minus sign for fermionic contributions.

The Wetterich equation, together with a given bare action in the formal \gls{UV} limit $k \to \infty$, constitutes an initial-value problem for the scale-dependent effective action $\Gamma_k$. 
Integrating the \gls{RG} flow into the \gls{IR} limit, $k \to 0$, yields the exact quantum effective action $\Gamma[\Phi]$, which contains all physical observables.
While $\Gamma_k$ depends on the choice of the regulator matrix $R_k$, the \gls{IR} limit is regulator-independent -- as long as the same bare action $S[\Phi]$ 
is used for initialization in the \gls{UV} limit~\cite{Wetterich:1992yh,Zorbach:2025drj}.

Since \cref{eq:wetterich-equation} is a highly complicated functional differential equation, which is in general not exactly solvable, one needs to employ so-called \textit{truncations}.
We define a truncation as a projection map ${\mathds{T}}$, which acts as an endomorphism on the space of all actions.
This map is inserted on the right-hand side of the Wetterich equation \eqref{eq:wetterich-equation} before the field derivatives are performed~\cite{Stoll:2025jor}, yielding the \textit{truncated Wetterich equation}
\begin{align*}
	\label{eq:truncated-wetterich-equation}\numberthis
	\partial_t \Gamma_k [ \Phi ] = \frac{1}{2}\,\mathrm{STr} \Big[  \big( ({\mathds{T}} \circ \Gamma_k)^{(2)} [ \Phi ] + R_k \big)^{-1} \cdot \partial_t R_k \Big] \, .
\end{align*}
Note that $\Gamma_k$ denotes a solution of the truncated Wetterich equation here and should not be confused with a solution of the exact Wetterich equation.
If ${\mathds{T}}$ is the identity, one recovers the exact Wetterich equation \eqref{eq:wetterich-equation}. 
The action ${\mathds{T}} \circ \Gamma_k$ is referred to as the \textit{truncated scale-dependent effective action} or truncation ``ansatz''.
The truncation map ${\mathds{T}}$ typically reduces the complexity of a given action $\Gamma_k$ by extracting a few (\gls{RG}-scale dependent) couplings/terms $\mathcal{O}_i(k)$ like the bosonic potential or wave-function renormalizations, which are then used to construct a new (much simpler) truncated action, ${\mathds{T}}\circ \Gamma_k$. 
This procedure leads to a much simpler and finite set of coupled ordinary/partial differential equations.
However, this reduction of information on the right-hand side of the flow equation may lead to systematic errors.
In general, solutions to the truncated Wetterich equation do not approximate solutions to the exact Wetterich equation unless the truncation is sufficiently accurate.
In general, there is no strict procedure to identify important couplings/terms like in perturbation theory. 
For example, in studies of critical phenomena in scalar field theories, the smallness of the anomalous dimension may be used to guide the construction of meaningful truncations, see, e.g., Ref.~\cite{Zorbach:2025drj} for a quantitative comparison of \gls{LPA} with lattice Monte Carlo simulations.
For an analysis of other phenomena, however, different approximation schemes may apply.
For a recent general discussion of systematic constructions of truncations, we refer to Ref.~\cite{Ihssen:2024miv}.

In this work, we are particularly interested in effective models incorporating quark and meson fields for in-medium \gls{QCD} and their treatment within the \gls{fRG} formalism, see, e.g., Refs.~\cite{Braun:2018svj,Braun:2009si,Oliveira:2011pg,Fu:2019hdw,Tripolt:2017zgc,Otto:2022jzl,Schaefer:2006ds,Ellwanger:1994wy,Jungnickel:1995fp,Berges:1997eu,Berges:1998sd,Zhang:2017icm,Ihssen:2023xlp,Murgana:2023pyx,Murgana:2025wsh}. 
Typically, such effective models lack \gls{UV} completeness, i.e., they do not exhibit a nontrivial \gls{UV} fixed point~\cite{Braun:2010tt,Gies:2010mqh,Braun:2012zq}.
Nevertheless, they can yield meaningful predictions within the \gls{fRG} formalism in the following way: 
Instead of initializing the scale-dependent effective action $\Gamma_k$ in the \gls{UV} limit with a bare action, we initialize it at some finite \gls{UV} cutoff (scale) $k=\Lambda$ with a model action, see, e.g., Refs.~\cite{Rennecke:2016tkm,Fu:2016tey,Herbst:2010rf,Zhang:2017icm}, which is the action \eqref{eq:bare_action} in this study.
The \gls{UV} parameters of this model action are tuned to reproduce a set of given \gls{IR} observables.
All remaining observables are then predictions of the effective model. 
It is important to note that the \gls{UV} cutoff is not uniquely defined, although it is often motivated as the scale where gluon dynamics effectively decouples from meson dynamics in \gls{fRG} calculations of \gls{QCD} in Landau gauge~\cite{Mitter:2014wpa,Braun:2014ata,Cyrol:2017ewj} due to the emergence of a gluon mass gap~\cite{Cornwall:1981zr,Ferreira:2025tzo,Aguilar:2021uwa,Oliveira:2011pg}.
Furthermore, the finiteness of the \gls{UV} cutoff entails that the corresponding initial-value problem has in general no simple path-integral analogue and consequently there is no regulator independence of the \gls{IR} limit of $\Gamma_k$ -- not even without truncations -- if $\Gamma_\Lambda$ is initialized with exactly the same model action for different regulators.
To achieve comparable results for different choices of the regulator and of the \gls{UV} cutoff, the \gls{UV} parameters of the model must instead be re-tuned, see, e.g., Refs.~\cite{Ihssen:2023xlp,Braun:2009si}. 
However, this re-tuning does not guarantee regulator- and \gls{UV} cutoff-independent predictions, not without truncations and in particular not with truncations~\cite{Ihssen:2023xlp}.

If we want to investigate effective models initialized at a finite \gls{UV} cutoff with the truncated Wetterich equation, we therefore face a central question: \textit{How sensitive are predictions to the choice of the regulator and \gls{UV} cutoff?}
A necessary condition to make meaningful predictions with an effective model in a certain truncation is that the dependence on the regulator and the \gls{UV} cutoff is weak, since the latter are at least to some degree arbitrary~\cite{Ihssen:2023xlp}. 
This property is referred to as \textit{\gls{RG} consistency}~\cite{Braun:2018svj}.
By \gls{UV}-cutoff dependence, we refer to changes in observables induced by variations of the \gls{UV} cutoff at fixed regulator, whereas regulator dependence denotes changes in observables induced by variations of the regulator at fixed \gls{UV} cutoff.
Note that a change of the regulator also alters the meaning of the actual values of the \gls{UV} cutoff (e.g., measured in units of low-energy observables), since both are inherently intertwined. 
However, in theories with a finite \gls{UV} cutoff, as is often the case for effective models, keeping the \gls{UV} cutoff fixed while varying the regulator provides a natural framework for studying regulator dependence.\footnote{
	By contrast, in theories for which the \gls{UV} cutoff can be removed, comparisons between different regulators are most meaningful in the corresponding continuum limit. 
	Note also that, in non-perturbative calculations, a regulator dependence in general persists even if the \gls{UV} cutoff can be removed.
}

For a variation of the \gls{UV} cutoff the change in an observable can be made quantitative: If a small relative change in the \gls{UV} cutoff $\Lambda \to \Lambda+\delta \Lambda$ induces a much smaller relative change in a scalar physical observable $O(\Lambda) \to O(\Lambda+\delta \Lambda)$, i.e., 
\begin{align*}
	\label{eq:RG-cons-cutoff}\numberthis
	\alpha=\Big|\frac{O(\Lambda+\delta \Lambda)-O(\Lambda)}{O(\Lambda)}\Big| \, \Big/ \, \Big|\frac{\delta \Lambda}{\Lambda}\Big| \ll 1 \, ,
\end{align*}
then this observable might legitimately be called \gls{RG}-consistent with respect to a variation of the \gls{UV} cutoff around $\Lambda$.
In the limit of $\delta \Lambda \to 0$, this relation becomes
\begin{align*}
	\label{eq:RG-cons-cutoff_infini}\numberthis
	\Big|\frac{\Lambda}{O(\Lambda)}  \partial_\Lambda O(\Lambda)\Big|  = \big| \partial_{\,\ln \Lambda} \ln O(\Lambda) \big| \ll 1 \, , 
\end{align*}
which is in accordance with the standard formulation of the \gls{RG}-consistency condition~\cite{Braun:2018svj}.
Note that \cref{eq:RG-cons-cutoff,eq:RG-cons-cutoff_infini} do not define a sharp threshold beyond which one may speak of \gls{RG} consistency.
A similar quantitative criterion is more difficult to formulate for variations of the regulator, but the principle remains: physical observables should not vary strongly under a change of the regulator. 
Note that the degree of \gls{RG} consistency may depend on the observable under consideration. 

In this work, we study the \gls{QMM} with a finite \gls{UV} cutoff and initialize it with the model action~\eqref{eq:bare_action} to investigate the \gls{RG} consistency of its predictions with particular focus on \gls{SSB} of the chiral symmetry.
More precisely, we study the formation of a nonzero chiral condensate associated with the symmetry-breaking pattern
\begin{align*}
	\label{eq:SSB-pattern}\numberthis
	\mathrm{SU}(2)_\mathrm{L} \times \mathrm{SU}(2)_\mathrm{R} \overset{\mathrm{SSB}}{\longrightarrow} \mathrm{SU}(2)_\mathrm{L+R}\,,
\end{align*}
where $\mathrm{SU}(2)_\mathrm{L+R}$ denotes the diagonal subgroup of the original chiral symmetry group~\cite{Schaefer:2006ds,Berges:1997eu}. 
To this end, we consider the scale-dependent effective potential defined by the projection prescription
\begin{align*}
	\label{eq:scale-dependent-effective-potential}\numberthis
	U_k(\bar{\sigma}) = \mathrm{proj}_U(\Gamma_k) = \frac{1}{V_d} \Gamma_k[\Phi=\Phi_0] \, ,
\end{align*} 
evaluated at the constant field configuration
\begin{align*}
	\label{eq:super-field-evaluation}\numberthis
	\Phi_0 = (\bar{\psi}=0,\psi=0,\phi=(\bar{\sigma},0,0,0))  \, .
\end{align*}
The global minimum $\bar{\sigma}_\mathrm{gs}$ of the effective potential $U = U_{k \to 0}$ is the chiral condensate and serves as the order parameter for chiral \gls{SSB}.
Note that, since we work with a nonzero explicit symmetry-breaking term in the action~\eqref{eq:bare_action} from here on, the order parameter cannot become exactly zero. 

\section{Framework}\label{sec:setup} 
\subsection{Local potential approximation and regulators}\label{sec:LPA}
In \gls{LPA}, the truncated scale-dependent effective action takes the form
\begin{align*}
	\label{eq:LPA-map} \numberthis
	{\mathds{T}} \circ \Gamma_k[\Phi] =
	\int_x \Big\{ \bar\psi \mathrm{i} \big[ &\slashed{\partial} -\mu \gamma_0 + h (\sigma + \mathrm{i} \vec\tau \cdot \vec\pi \gamma_\mathrm{ch})\big] \psi \\
	&+ \tfrac{1}{2} \phi \cdot (-\partial^2) \phi + U_k(\phi)\Big\} \,,
\end{align*}
which implies that the scale-dependent effective potential~$U_k$ is the only contribution from $\Gamma_k$ that dynamically feeds back into the right-hand side of the flow equation. 
This approximation thus incorporates bosonic fluctuations through the meson kinetic term and the scale-dependent effective potential.

We restrict ourselves to three-dimensional regulators that act only on the spatial momenta (see, e.g., Refs.~\cite{Litim:2006ag,Blaizot:2006rj}), leaving the temporal direction unregularized.
In the quark sector, the regulator term reads
\begin{align*}
	\label{eq:std_reg_class_fermi}\numberthis
	\Delta S^\mathrm{q}_k &= - \frac{1}{\beta} \sum_{p_{0} \in \tilde{\Omega}_\mathrm{q}} \int_{\mathds{R}^{d-1}} \frac{\d^{d-1} \vec{p}\,}{(2\uppi)^{d-1}} \, \tilde{\bar{\psi}}_p \, \slashed{\vec{p}} \, \tilde r\left(\frac{\vec{p}^{\,2}}{k^2}\right) \tilde{\psi}_p \, ,
\end{align*}
with fermionic Matsubara frequencies $\tilde\Omega_\mathrm{q} = \{\frac{2 \uppi}{\beta} (n+\frac{1}{2}) \; \vert \; n\in \mathds{Z}\}$.
In the meson sector, the regulator term is given by
\begin{align*}
	\label{eq:std_reg_class_bose}\numberthis
	\Delta S^\mathrm{m}_{k} &= \frac{1}{\beta} \sum_{p_{0} \in \tilde{\Omega}_\mathrm{m}} \int_{\mathds{R}^{d-1}} \frac{\d^{d-1} \vec{p}\,}{(2\uppi)^{d-1} } \, \tilde{\phi}_{p} \frac{\vec{p}^{\,2}}{2} r\left(\frac{\vec{p}^{\,2}}{k^2}\right) \tilde{\phi}_{-p}\, ,
\end{align*}
with bosonic Matsubara frequencies $\tilde\Omega_\mathrm{m} = \{\frac{2 \uppi}{\beta} n\; \vert \; n \in \mathds{Z}\}$.
The regulator matrix $R_k$ in \cref{eq:wetterich-equation} is the Hessian of the combined quark and meson regulator terms with derivatives taken with respect to the components of $\Phi$.

The regulator terms in \cref{eq:std_reg_class_fermi,eq:std_reg_class_bose} are specified by the regulator shape functions $\tilde{r}(z)$ in the quark sector and $r(z)$ in the meson sector. 
We relate them via 
\begin{align*}
	\numberthis\label{eq:regulator-shape-functions-connection}
	\tilde r(z) = \sqrt{r(z)+1}-1 \, ,
\end{align*}
which is a natural choice, as becomes apparent in the energy functions of the flow equation for the scale-dependent effective potential, see below.
The meson regulator shape function $r(z)$ is required to satisfy the standard conditions
\begin{align*}
	\numberthis\label{eq:regulator-shape-function-constraints}
	\lim_{z\to 0} z\,r(z) > 0 \, , \quad \lim_{z\to \infty} r(z) = 0 \, ,
\end{align*}
see, e.g., Ref.~\cite{Wetterich:1992yh,Litim:2000ci}, which implicitly constrain the admissible forms of $\tilde r(z)$ as well.
The first condition ensures that the regulator term diverges in the \gls{UV} limit. 
Additionally, it prevents \gls{IR} divergences in the momentum integrals of the flow equation for the scale-dependent effective potential by effectively inserting a scale- and momentum-dependent mass term into the propagator.
The second condition guarantees that the regulator term is removed in the \gls{IR} limit.

In the following we employ three different regulator shape functions: the Litim regulator shape function~\cite{Litim:2000ci,Litim:2001up}, an exponential regulator shape function (Exp2), and a smooth Litim-like regulator shape function (SL), where the latter two were introduced in Ref.~\cite{Zorbach:2024zjx}:
\begin{subequations}
	\label{eq:shape_func}
	\begin{align}
		r_{\mathrm{Litim}}(z) &= \left(\frac{1}{z} - 1\right) \Theta(1-z) \, , \\
		r_{\mathrm{Exp2}}(z) &= \frac{1}{1-\mathrm{e}^{-z-z^2}} - 1 \, , \\
		r_{\mathrm{SL}}(z) &= \exp(-\frac{1}{z-\frac{1}{2}}) \Theta \left(z-\frac{1}{2}\right) + \frac{1}{z} -1 \, .
	\end{align}
\end{subequations} 
All three regulator shape functions satisfy the constraints~\eqref{eq:regulator-shape-function-constraints}.
Our choice of regulator shape functions is distinguished by its diversity in terms of their properties.
To be specific, it includes the widely used but non-differentiable three-dimensional Litim regulator shape function, which allows for an analytic evaluation of the momentum integrals in the flow equation of the scale-dependent effective potential and thereby considerably simplifies calculations.
Moreover, we include a representative from the class of the analytic exponential regulator shape functions, $r_{\mathrm{Exp2}}(z)$.
Finally, we consider $r_{\mathrm{SL}}(z)$, which is smooth despite being defined piecewise via a bump function.
Thus, these regulator shape functions differ strongly in their analytic properties.
Moreover, with respect to the principle of strongest singularity -- a recently introduced optimization criterion, which provides a mathematical order relation for regulators -- the Litim and Exp2 regulator shape functions are located at opposite ends of the spectrum, with the Litim regulator being optimal in \gls{LPA} in the case of three- and four-dimensional scalar field theories~\cite{Zorbach:2024zjx}.\footnote{The principle of strongest singularity allows to order regulators according to how fast the effective potential becomes flat/convex in the \gls{RG} flow.}
Note that, at any nonzero temperature, dimensional reduction sets in in the \gls{IR} limit, such that the \gls{QMM} effectively reduces to a three-dimensional scalar field theory for which the Litim regulator shape function can be considered optimal according to the principle of strongest singularity~\cite{Zorbach:2024zjx} and also other optimization criteria~\cite{Litim:2000ci,Litim:2001up,Litim:2001fd,Pawlowski:2005xe}.
Of course, it is not immediately clear that the three-dimensional Litim regulator is also optimal with respect to, e.g., the principle of strongest singularity in the case of a model where the \gls{RG} flow ranges from an effectively four-dimensional \gls{UV} limit with scalar and fermion fields down to an \gls{IR} limit effectively described by a three-dimensional scalar field theory. 
In any case, given the fact that the regulator shape functions considered here differ in several essential aspects, we expect them to provide a suitable basis for exploring regulator dependences.

As stated above, our selection of regulator shape functions has been guided by the principle of strongest singularity since it equips us with a simple order relation for regulators~\cite{Zorbach:2024zjx}.
We add that other optimization criteria for regulators exist in the literature.
For example, an early discussion of the optimization of \gls{RG} flows based on the principle of minimum sensitivity~\cite{Stevenson:1981vj} can be found in Ref.~\cite{Liao:1999sh}. 
This principle and the corresponding optimization criterion have been employed in \gls{fRG} computations of critical exponents and universal amplitude ratios of scalar field theories in, e.g., Refs.~\cite{Canet:2002gs,Canet:2003qd,Balog:2019rrg,DePolsi:2020pjk,DePolsi:2021cmi,DePolsi:2022wyb}.
Optimization based on a maximization of the gap in the regularized two-point function in the \gls{RG} flow has been discussed in Refs.~\cite{Litim:2000ci,Litim:2001up,Litim:2001fd}. 
Note that this optimization criterion includes the principle of minimum sensitivity as a special case.
Functional optimization as introduced in Ref.~\cite{Pawlowski:2005xe} and further discussed in Ref.~\cite{Pawlowski:2015mlf} is built on supplementing the aforementioned maximization of the gap in the propagator by the minimization of the full (regularized) two-point function entering the Wetterich equation and allows to optimize regulators at any given order of the derivative expansion.
At least in \gls{LPA}, functional optimization and optimization according to the principle of strongest singularity single out the same regulator for covariant theories in the vacuum limit, i.e., the Litim regulator~\cite{Litim:2000ci,Litim:2001up,Litim:2001fd}.
Independently of that, a criterion for optimization based on constraints stemming from conformal invariance has recently been proposed~\cite{Balog:2020fyt,Delamotte:2024xhn}.

The regulator dependence in \gls{QMM}-type models has also been addressed in other recent \gls{RG} studies, such as Refs.~\cite{Otto:2022jzl,Ihssen:2023xlp}. 
Whereas the  impact of mass-like regulators within \gls{LPA} is discussed in Ref.~\cite{Otto:2022jzl}, the effect of different relative \gls{IR}-cutoff scales in the quark and meson regulators on physical observables in a self-consistent \gls{LPA} framework is analyzed in Ref.~\cite{Ihssen:2023xlp}.
In our present work, however, the focus is on a consistent parameter-fixing procedure and the resulting systematic analysis of the variation of physical observables under a variation of momentum regulators and the \gls{UV} cutoff in \gls{LPA}, which underlies many phenomenological studies of the QCD phase diagram.

\subsection{Flow equation}\label{sec:flow-eq}
With the truncation and regulators at hand, the flow equation for the scale-dependent effective potential with a general regulator shape function can be derived and takes the form 
\begin{widetext}
	\begin{align*}
		\label{eq:flow-eq}\numberthis
		\partial_t U_k(\bar{\sigma}) = &-d_\gamma N_\mathrm{c} N_\mathrm{f} \int_{\mathds{R}^{d-1}} \frac{ \d^{d-1} \vec{p}}{(2\uppi)^{d-1}} \, 2 \frac{\vec{p}^{\,2}}{k^2} \tilde{r}^\prime(\vec{p}^{\,2}/k^2) \vec{p}^{\,2} \left[1+\tilde{r}(\vec{p}^{\,2}/k^2)\right]  
		\frac{1}{2 E_\mathrm{q}} \bigl[ 1 - n_\mathrm{F}(\beta (E_\mathrm{q}+\mu)) -n_\mathrm{F}(\beta (E_\mathrm{q}-\mu)) \bigr]  \\
		&+ \int_{\mathds{R}^{d-1}} \frac{ \d^{d-1} \vec{p}}{(2\uppi)^{d-1}} \, \vec{p}^{\,2} \, \frac{\vec{p}^{\,2}}{k^2} r^\prime(\vec{p}^{\,2}/k^2)
        \biggl\{ \frac{1}{2 E_\sigma} \bigl[ 1+ 2 n_\mathrm{B}(\beta E_\sigma) \bigr] + (N_\mathrm{m} -1) \frac{1}{2 E_\pi} \bigl[ 1+ 2 n_\mathrm{B}(\beta E_\pi) \bigr] \biggr\} \, .
	\end{align*}
\end{widetext}

The Fermi-Dirac distribution function $n_\mathrm{F}(a)=1/(\mathrm{e}^a+1)$ and Bose-Einstein distribution function $n_\mathrm{B}(a)=1/(\mathrm{e}^a-1)$ result from an analytic evaluation of the Matsubara sums that occur in the derivation of this equation.
The auxiliary energy functions entering \cref{eq:flow-eq} are given by 
\begin{subequations}
	\label{eq:energy-funcs}
	\begin{align}
		E_\mathrm{q} &= \sqrt{\vec{p}^{\,2}\bigl[1+\tilde{r}(\vec{p}^{\,2}/k^2)\bigr]^2 + M^2_\mathrm{q}} \, ,\\
		E_\sigma &= \sqrt{ \vec{p}^{\,2} \bigl[ 1+ r(\vec{p}^{\,2}/k^2) \bigr] + M^2_{\sigma}} \, , \\
		E_\pi &= \sqrt{\vec{p}^{\,2} \bigl[ 1+ r(\vec{p}^{\,2}/k^2) \bigr] + M^2_{\pi}} \, ,
	\end{align}
\end{subequations}
where 
\begin{subequations}
	\label{eq:mass-funcs}
	\begin{align}
		M^2_\mathrm{q}(\bar{\sigma}) &= (h \bar{\sigma})^2 \, ,\\
		M^2_{\sigma}(k,\bar{\sigma}) &= \partial^2_{\bar{\sigma}} U_k(\bar{\sigma}) \, ,\\
		M^2_{\pi}(k,\bar{\sigma}) &= [\partial_{\bar{\sigma}}U_k(\bar{\sigma}) + c]/\bar{\sigma} \, ,
	\end{align}
\end{subequations}
are the squared mass functions.
Note that, using the relation~\eqref{eq:regulator-shape-functions-connection} between $r$ and $\tilde r$, i.e., $[1+\tilde r(z)]^2 = 1 + r(z)$, the kinetic contributions to the energy functions coincide, i.e., the quark and meson contributions to the flow equation~\eqref{eq:flow-eq} are regulated consistently.
In the following, we use the terms regulator and regulator shape function interchangeably.

The flow equation~\eqref{eq:flow-eq} is a nonlinear partial differential equation that requires proper numerical treatment.
This is achieved by reformulating it as a flow equation for $\partial_{\bar{\sigma}} U_k$, thus bringing it into conservative form~\cite{Grossi:2019urj,Koenigstein:2021syz}. 
This allows us to apply methods borrowed from numerical fluid dynamics, see, e.g., Refs.~\cite{Stoll:2021ori,Zorbach:2024zjx,Rais:2024jio,Zorbach:2024rre,Zorbach:2025drj,Stoll:2025jor,Koenigstein:2025sse,Koenigstein:2021syz,Koenigstein:2021rxj,Steil:2021cbu,Murgana:2023xrq,Jeong:2024rst} for applications. 
In our case, we employ the Kurganov-Tadmor scheme~\cite{Kurganov:2000ovy}, which is designed to handle advection-diffusion-type partial differential equations. 
For details and numerical parameters, see \cref{app:num}.

\subsection{Parameter fixing and UV-cutoff choice}\label{sec:param-fix}
After introducing the flow equation for general three-dimensional regulators, we now turn to the parameter-fixing procedure for the different regulators and \gls{UV} cutoffs $\Lambda$ that are necessary to analyze \gls{RG} consistency in  \gls{QMM} studies.
From the model action \eqref{eq:bare_action} we infer the initial condition for the scale-dependent effective potential:
\begin{align*}
	\label{eq:init-cond}\numberthis
	U_\Lambda(\bar{\sigma}) \;=\; \frac{m^2}{2}\, \bar{\sigma}^2 \;+\; \frac{\lambda}{4}\, \bar{\sigma}^4 \;-\; c \,\bar{\sigma} \, .
\end{align*}
For the scale~$\Lambda$, we consider the set $\Lambda \in \{0.6\GeV,0.8\GeV,1.0\GeV,1.2\GeV\}$, which lies in the range of \gls{UV} cutoffs typically employed in \gls{QMM} studies in the literature, see, e.g., Refs.~\cite{Herbst:2010rf,Fu:2016tey,Rennecke:2016tkm,Tripolt:2017zgc,Zhang:2017icm,Eser:2018jqo}.
The parameter-fixing procedure now aims to tune the free parameters $m^2$, $\lambda$, and $c$ in the initial condition, as well as the Yukawa coupling $h$, which enters the flow equation for the scale-dependent effective potential, such that we recover the values of a given set of vacuum low-energy observables from the \gls{RG} flow in the \gls{IR}.
Due to the regulator and \gls{UV}-cutoff dependence discussed in \cref{sec:FRGandEFT}, it is not possible to find a single set of \gls{UV} parameters and use it consistently for different regulator and \gls{UV}-cutoff combinations.
To ensure comparability of our results, we rather have to fix the initial condition separately for each combination such that the same physical observables are reproduced in the vacuum at the \gls{IR} scale $k_{\mathrm{IR}} = 50\,\MeV$.
In our context, the ``vacuum'' is considered to be at $\mu = 0$ and $T = 0.0001\,\text{GeV}$, which we assume to be a sufficiently accurate approximation of the zero-temperature limit.
Four \gls{IR} observables are employed for the parameter fixing.
Specifically, we consider the vacuum minimum $\bar{\sigma}_\mathrm{gs}^\mathrm{vac}$ of the effective potential, which can be identified with the pion decay constant~$f_{\pi}$; the vacuum curvature mass $m_\mathrm{c}^\mathrm{vac}$ of the sigma meson, defined as the square root of the curvature of the effective potential at its minimum; the vacuum constituent quark mass~$m_{\text{q}}^{\text{vac}}:=h \bar{\sigma}_\mathrm{gs}^\mathrm{vac}$; and the vacuum curvature mass of the pion determined by $(m_{\pi}^{\text{vac}})^2:= c/\bar{\sigma}_\mathrm{gs}^\mathrm{vac}$.
The latter two observables are used to fix the Yukawa coupling $h$ and the explicit symmetry-breaking parameter $c$.
Note that the term~$c \bar{\sigma}$ in the scale-dependent effective potential remains unchanged under the \gls{RG} flow.
The Yukawa coupling is generally scale-dependent also in \gls{LPA}, but we neglect its \gls{RG} flow in our study.
This leaves us with a set of two \gls{UV} parameters~$(\lambda, m^2)$ that have to be tuned to reproduce the desired \gls{IR} values of the minimum $\bar{\sigma}_\mathrm{gs}^\mathrm{vac}$ and the curvature mass $m_\mathrm{c}^\mathrm{vac}$ of the sigma meson. 
The \gls{UV} parameter sets $(\lambda, m^2)$ obtained for all \gls{UV}-cutoff and regulator combinations in the vacuum are then used for the \gls{RG} flows at nonzero temperature $T$ and chemical potential $\mu$, in order to calculate the corresponding in-medium observables.

\begin{figure}
	\includegraphics{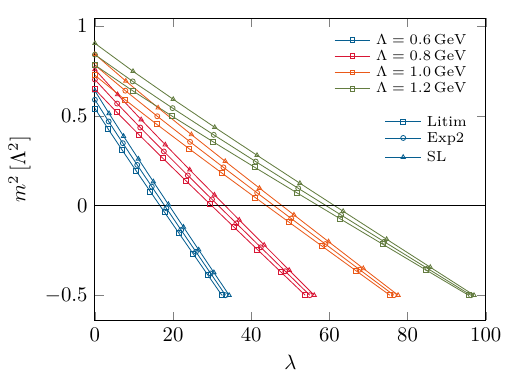}
	\caption{\label{fig:lam_vs_m2} The value of $m^2$ as a function of $\lambda$ for each regulator and \gls{UV} cutoff combination considered in this work. All $(\lambda,m^2)$ values on these curves lead to $\bar{\sigma}_\mathrm{gs}^\mathrm{vac} = 0.093 \GeV$.}
\end{figure}

In the remainder of this section, we describe in detail how the parameter-fixing procedure is implemented.
It begins by imposing $f_\pi = 0.093 \GeV$, $m_{\text{q}}^{\text{vac}}\approx 0.317\GeV$, and~$m_{\pi}^{\text{vac}}\approx 0.135\GeV$.
This corresponds to choosing $c=0.001695\GeV^3$, $h=3.41$ and $\bar{\sigma}_\mathrm{gs}^\mathrm{vac}=0.093 \GeV$, which restricts the admissible $(\lambda,m^2)$ values for each regulator and \gls{UV}-cutoff combination to a one-dimensional curve in the $(\lambda,m^2)$-plane, see~\cref{fig:lam_vs_m2}. 
The vacuum curvature mass $m_\mathrm{c}^\mathrm{vac}$ of the sigma meson, however, varies along these curves. 
Interestingly, the curves exhibit an approximately linear relation between $\lambda$ and $m^2$.
This is a nontrivial observation, but can be partially understood as follows:  
If we retain only the quark contribution, the flow equation~\eqref{eq:flow-eq} reduces to an ordinary differential equation, which corresponds to the mean-field approximation, see, e.g., Ref.~\cite{Stoll:2021ori}. 
Together with the initial condition~\eqref{eq:init-cond} we then obtain
\begin{align*}
	\label{eq:MF-pot}\numberthis
	0&\equiv \partial_{\bar{\sigma}} U(\bar{\sigma}_\mathrm{gs}^\mathrm{vac})\\
	&= m^2 \bar{\sigma}_\mathrm{gs}^\mathrm{vac} + \lambda (\bar{\sigma}_\mathrm{gs}^\mathrm{vac})^3 - c + \int_{k_\mathrm{IR}}^{\Lambda} \frac{\d k}{k} \partial_t \partial_{\bar{\sigma}} U_k(\bar{\sigma}_\mathrm{gs}^\mathrm{vac})\, ,
\end{align*}
at the \gls{IR} minimum. 
By solving this expression for $m^2/\Lambda^2$, we obtain a linear dependence on $\lambda$, with slope $-(\bar{\sigma}_\mathrm{gs}^\mathrm{vac})^2/\Lambda^2$ and a positive intercept at $\lambda=0$. 
While this reasoning qualitatively captures the trend shown in \cref{fig:lam_vs_m2}, quantitative deviations arise. 
They result from the meson contributions, which tend to restore the chiral symmetry. 
Thus, smaller values of $m^2$ and/or $\lambda$ are required to keep $\bar{\sigma}_\mathrm{gs}^\mathrm{vac}=f_\pi$ fixed.
In particular, for large values of $\lambda$, the meson contributions become significant already at the start of the \gls{RG} flow, requiring much lower values of $m^2$. 
In principle, the meson contributions could spoil the linear relation derived from the pure quark contribution, but, as seen in \cref{fig:lam_vs_m2}, linearity approximately persists even in \gls{LPA}.
At least for the Litim regulator shape function, this observation may at least partially be traced back to the fact that, at the initial \gls{RG} scale, the meson terms in \cref{eq:flow-eq} at $\bar{\sigma}=\bar{\sigma}_\mathrm{gs}^\mathrm{vac}$ depend approximately linearly on $\lambda$, for sufficiently small $\lambda$.

In \cref{fig:lam_vs_m2}, we furthermore observe that the dependence of the \gls{UV} parameters on the regulator is weaker than their dependence on the \gls{UV} cutoff within the considered \gls{UV} cutoff range.
Note that there are canonical bounds on the extension of these curves: 
First, the requirement to have a scale-dependent effective potential which is bounded from below enforces $\lambda>0$. 
Second, singularities in the meson propagators must be avoided at the \gls{UV}-cutoff scale, i.e., the energy functions~\eqref{eq:energy-funcs} must remain positive real-valued.
This approximately translates into $m^2 \gtrsim -\Lambda^2$, which can best be seen from the flow equation \eqref{eq:flow-eq} with the Litim regulator.
In \cref{fig:lam_vs_m2}, however, we restrict ourselves to $m^2 > -0.5\Lambda^2$ to ensure numerical stability for all regulator choices.

Next, we fix the remaining freedom by imposing a value for the vacuum curvature mass $m_\mathrm{c}^\mathrm{vac}$ of the sigma meson. 
However, not all values of $m_\mathrm{c}^\mathrm{vac}$ are accessible from the curves in \cref{fig:lam_vs_m2}.
\begin{figure}
	\includegraphics{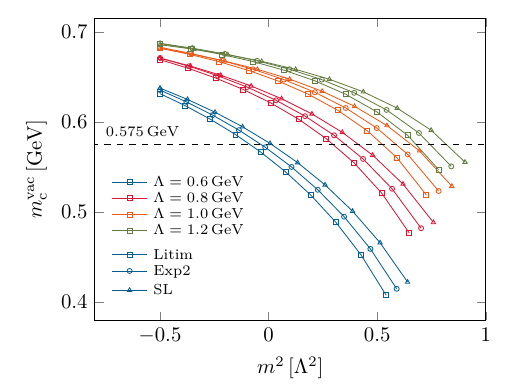}
	\caption{\label{fig:m2_vs_mIR} Vacuum curvature mass of the sigma meson~$m^{\text{vac}}_{\text{c}}=m^{\text{vac}}_{\text{c}}(\lambda,m^2)$ as a function of $m^2$ along the lines in the $(\lambda,m^2)$ plane given in \cref{fig:lam_vs_m2}.}
\end{figure}
To determine the admissible range, we evaluate $m_\mathrm{c}^\mathrm{vac}$ for all $(\lambda,m^2)$ values on the curves and parametrize it with $m^2$. 
This again yields one-dimensional curves in the $(m^2,m_\mathrm{c}^\mathrm{vac})$-plane for each regulator and \gls{UV} cutoff, see \cref{fig:m2_vs_mIR}.\footnote{Note that the curvature mass of the sigma meson is only a meaningful tuning quantity in the presence of explicit symmetry breaking. In the chiral limit it vanishes logarithmically in the \gls{IR} limit.}
Interestingly, we find that the range of attainable $m_\mathrm{c}^\mathrm{vac}$ values shrinks with increasing \gls{UV} cutoff. 
A discussion of this observation will be presented elsewhere~\cite{forthcomingpaper}. 
In any case, from \cref{fig:m2_vs_mIR} we infer that the interval $m_\mathrm{c}^\mathrm{vac} \in [0.56\GeV,0.63\GeV]$ is simultaneously accessible for all considered regulators and \gls{UV} cutoffs.
In the following, we choose $m_\mathrm{c}^\mathrm{vac}=0.575\GeV$ from this interval.
The corresponding $m^2$ values are determined from the intersection of the horizontal line at $m_\mathrm{c}^\mathrm{vac}=0.575\GeV$ with the curves in \cref{fig:m2_vs_mIR}. 
Using these values in \cref{fig:lam_vs_m2} then yields the corresponding $\lambda$ values. 
The resulting \gls{UV} parameters are listed explicitly in \cref{tab:UV-param}.
This procedure guarantees that, for each regulator and \gls{UV}-cutoff combination, we have $\bar{\sigma}_\mathrm{gs}^\mathrm{vac} = 0.093 \GeV$ and $m_\mathrm{c}^\mathrm{vac}=0.575\GeV$ in the \gls{IR}.

Finally, let us comment on the choice of the \gls{IR} scale $k_{\mathrm{IR}}$, at which the observables are evaluated. In principle, $k_{\mathrm{IR}}$ should be taken to zero, but this is not feasible within our numerical framework. 
Instead, we select a small but nonzero value such that the observables of interest remain essentially insensitive to variations of $k_{\mathrm{IR}}$.
In the vacuum, this insensitivity sets in once the masses $m_\mathrm{q}^\mathrm{vac}$, $m_\mathrm{c}^\mathrm{vac}$, and $m_\pi^\mathrm{vac}$ dominate the energy functions \eqref{eq:energy-funcs}, i.e., when $k_{\mathrm{IR}} \ll m_{\pi}^\mathrm{vac},\, m_\mathrm{q}^\mathrm{vac},\, m_\mathrm{c}^\mathrm{vac}$. Therefore, in our study $k_{\mathrm{IR}} = 50\,\mathrm{MeV}$ is a suitable choice in the vacuum.
\begin{table*}[ht]
    \centering
    \setlength\extrarowheight{8pt}
    \begin{ruledtabular}
        \begin{tabular}{c c c c}
            & Litim & Exp2 & SL \\
            $\Lambda=0.6\GeV$ & $m^2=-0.03006\GeV^2, \; \lambda=19.260$ & $m^2=-0.01188\GeV^2, \; \lambda=18.840$ & $m^2=0.00551\GeV^2, \; \lambda=18.534$ \\
            $\Lambda=0.8\GeV$ & $m^2=0.19183\GeV^2, \; \lambda=15.697$ & $m^2=0.22897\GeV^2, \; \lambda=15.080$ & $m^2=0.26780\GeV^2, \; \lambda=14.497$ \\
            $\Lambda=1.0\GeV$ & $m^2=0.52987\GeV^2, \; \lambda=11.230$ & $m^2=0.59272\GeV^2, \; \lambda=10.490$ & $m^2=0.66198\GeV^2, \; \lambda=9.670$ \\
            $\Lambda=1.2\GeV$ & $m^2=0.98611\GeV^2, \; \lambda=6.561$ & $m^2=1.08151\GeV^2, \; \lambda=5.773$ & $m^2=1.19017\GeV^2, \; \lambda=4.790$ \\
        \end{tabular}
    \end{ruledtabular}
    \caption{\label{tab:UV-param} Results of the parameter-fixing procedure for $m^2$ and $\lambda$ for the different \gls{UV}-cutoff and regulator combinations.
	All \gls{UV}-parameter sets reproduce $\bar{\sigma}_\mathrm{gs}^\mathrm{vac} = 0.093 \GeV$ and $m_\mathrm{c}^\mathrm{vac}=0.575\GeV$ in the vacuum for their respective \gls{UV} cutoff and regulator.}
\end{table*}
\section{RG-consistency of the phase structure of the quark-meson model} \label{sec:results}
In this section, we study \gls{RG} consistency of the phase structure of the \gls{QMM}. 
To this end, we investigate the effective potential and related observables computed with different \gls{UV} cutoffs and regulators at various values of $\mu$ and $T$, based on the parameter-fixing procedure introduced in~\cref{sec:param-fix}.
We will use $k_\mathrm{IR}=50\MeV$ for our computations of the effective potential and will comment on the reliability of this choice away from the vacuum limit later.

\begin{figure*}
	\subfloat[\label{fig:sigma_vs_U_vac}%
	]{%
		\includegraphics{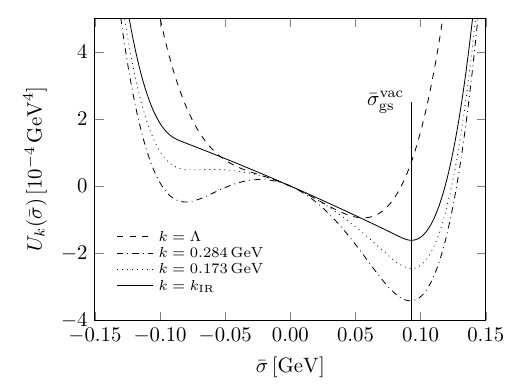}%
	}\hfill
	\subfloat[\label{fig:sigma_vs_u_vac}%
	]{%
		\includegraphics{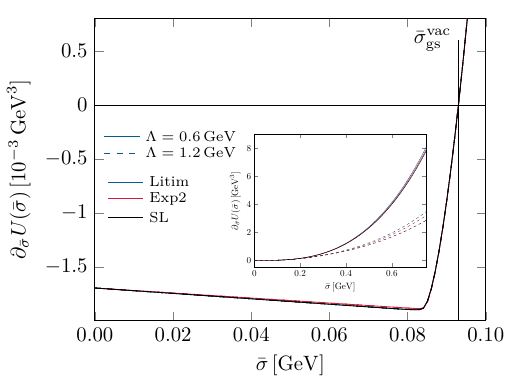}%
	}
	\caption{\label{fig:vac} Left: The scale-dependent effective potential for the Litim regulator with $\Lambda=0.6 \GeV$ is displayed for various \gls{RG} scales at $\mu=0$ and $T=0.0001\GeV$. 
	Right: The derivative of the effective potential $\partial_{\bar{\sigma}} U_{k=k_\mathrm{IR}}$ at $\mu=0$ and $T=0.0001\GeV$, but for various regulator and \gls{UV}-cutoff combinations. The inset in the right panel illustrates the global behavior of $\partial_{\bar{\sigma}} U_{k=k_\mathrm{IR}}$. In both panels, the vertical line indicates the position of the \gls{IR} value of~$\bar{\sigma}_\mathrm{gs}^\mathrm{vac}$, i.e., the minimum of $U=U_{k_\mathrm{IR}}$ or equivalently the zero of $\partial_{\bar{\sigma}} U$.}
\end{figure*}

We begin by exemplarily showing the scale dependence of the effective potential~$U_k$ (up to a constant) as obtained for the Litim regulator and $\Lambda=0.6\GeV$ in the vacuum for various \gls{RG} scales in \cref{fig:sigma_vs_U_vac}.
We observe that the contribution from the quarks to the flow equation deepens the minimum during the \gls{RG} flow, while the bosonic fluctuations render the potential convex.
Due to the explicit symmetry breaking, the effective potential is tilted to the right and exhibits a pronounced negative slope rather than a flat region as it would be the case for $c=0$. 
Eventually, the minimum of the effective potential is located at $\bar{\sigma}^{\mathrm{vac}}_\mathrm{gs} = 0.093\GeV$, in agreement with our parameter fixing.

In the following, however, we consider the first derivative of the effective potential, $\partial_{\bar{\sigma}} U$ (where $U\equiv U_{k=k_\mathrm{IR}}$), since it is the direct numerical output of our approach and comprises the full information of the effective potential up to a constant. 
The latter is of no relevance for the present work.
In \cref{fig:sigma_vs_u_vac}, the \gls{IR} result $\partial_{\bar{\sigma}} U$ is shown in the vacuum, for all combinations of the three regulators Litim, Exp2, SL and for the considered minimal and maximal values for the  \gls{UV} cutoff, $\Lambda=0.6\GeV$ and $\Lambda=1.2\GeV$, computed with their respective \gls{UV} parameters, see \cref{tab:UV-param}. 
The explicit symmetry breaking term translates into a constant downward shift of $\partial_{\bar{\sigma}} U$ by $-c$.
The zero of $\partial_{\bar{\sigma}} U$ is found at $\bar{\sigma}_\mathrm{gs}^\mathrm{vac} = 0.093\GeV$ consistent with the minimum of $U$.
We find very good agreement of $\partial_{\bar{\sigma}} U$ for the different regulators and \gls{UV} cutoffs in \cref{fig:sigma_vs_u_vac} both around $\bar{\sigma}_\mathrm{gs}^\mathrm{vac}$, which is of course trivial because of the parameter-fixing procedure, but also at smaller values of the field~$\bar{\sigma}$, where $\partial_{\bar{\sigma}} U$ becomes flat. 
The latter agreement is nontrivial, and it is only for the Exp2 regulator that we observe small but visible deviations.
In addition, the inset in \cref{fig:sigma_vs_u_vac} shows the global behavior of $\partial_{\bar{\sigma}} U$, i.e., the regime associated with large values of the field~$\bar{\sigma}$.
There, we observe strong deviations for the different \gls{UV}-cutoff and regulator combinations.
This is expected, since the large-field asymptotics of the initial condition \eqref{eq:init-cond} differs for the various \gls{UV}-parameter sets and is not modified in the \gls{RG} flow. 
In fact, quantum fluctuations are suppressed in this regime associated with~$\bar{\sigma} \gg \Lambda$.
Thus, differences persist even in the \gls{IR}.
This observation is consistent with our discussion in~\cref{sec:FRGandEFT}: re-tuning a finite set of parameters is not expected to yield the same results for all observables computed using different regulators and \gls{UV} cutoffs, i.e., not all the derivatives of the effective potential at $\bar{\sigma}_\mathrm{gs}^\mathrm{vac}$ as obtained using different regulators and \gls{UV} cutoffs will in general agree.
This is also reflected in the deviations in the global structure of the effective potentials.

We now turn to the case of nonzero $\mu$ and $T$.
For small values of $\mu$ and $T$, the invariance of $\partial_{\bar{\sigma}} U$ around its zero under variations of the \gls{UV} cutoff and the regulator persists, indicating \gls{RG} consistency in this regime.
At larger $\mu$ and $T$, however, artifacts from the \gls{UV} cutoff and the regulator inevitably appear, since our vacuum parameter-fixing procedure does not account for $\mu$- and $T$-induced differences of \gls{RG} flows arising between different regulators and \gls{UV} cutoffs. 
In the following, we investigate the regions in $(\mu,T)$ where such artifacts become significant and assess whether \gls{UV}-cutoff effects or regulator effects are dominant.
\begin{figure*}
	\subfloat[\label{fig:sigma_vs_u_smaller_T}%
	]{%
		\includegraphics{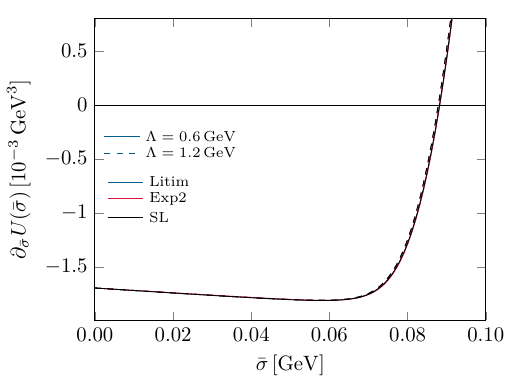}%
	}\hfill
	\subfloat[\label{fig:sigma_vs_u_great_T}%
	]{%
		\includegraphics{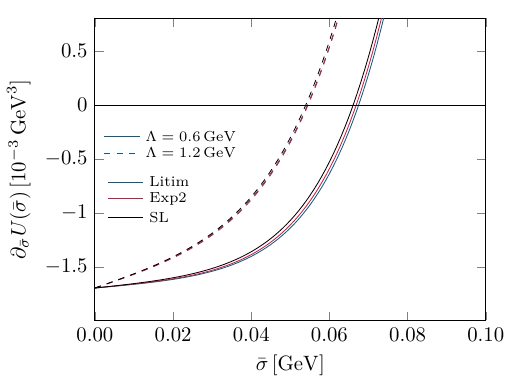}%
	}
	\caption{\label{fig:sigma_vs_u_high_T} Derivative of the effective potential at $\mu=0$ and $T=0.1\GeV$ (left panel) and $T=0.17\GeV$ (right panel) for different regulator and \gls{UV}-cutoff combinations.}
\end{figure*}

In \cref{fig:sigma_vs_u_high_T} we show $\partial_{\bar{\sigma}} U$ at $\mu=0$ for two temperatures $T=0.1\GeV$ and $T=0.17\GeV$, again for all combinations of the three regulators Litim, Exp2 and SL, and for the considered minimum and maximum \gls{UV} cutoffs $\Lambda=0.6\GeV$ and $\Lambda=1.2\GeV$. 
First we observe that $\bar{\sigma}_\mathrm{gs}$ decreases continuously with increasing temperature for all regulators and \gls{UV} cutoffs. 
This is the remnant of the second-order phase transition in this region of the phase diagram which becomes a crossover transition here due to the explicit symmetry breaking. 
Comparing \cref{fig:sigma_vs_u_smaller_T,fig:sigma_vs_u_great_T} shows that \gls{UV}-cutoff and regulator dependences grow with increasing temperature.
We generally find that \gls{UV}-cutoff effects within the considered \gls{UV}-cutoff range are significantly stronger than regulator effects.
In general, a larger \gls{UV} cutoff leads to a smaller value of $\bar{\sigma}_\mathrm{gs}$ at a given nonzero temperature.
Note that we have verified that $\partial_{\bar{\sigma}} U$ near $\bar{\sigma}_\mathrm{gs}$ is stable with respect to variations of $k_\mathrm{IR}$. 

As discussed in \cref{sec:FRGandEFT}, we can quantify the \gls{UV}-cutoff dependence of, e.g., the minimum $\bar{\sigma}_\mathrm{gs}$. 
To this end, we calculate $\alpha$ from \cref{eq:RG-cons-cutoff} for $O=\bar{\sigma}_\mathrm{gs}$, as an example for the Litim regulator.
More precisely, we use the minima obtained for the \gls{UV} cutoffs $\Lambda \in \{0.6\GeV,0.8\GeV,1.0\GeV,1.2\GeV\}$ and determine $\alpha$ for adjacent values of $\Lambda$. 
The difference $\delta \Lambda=0.2\GeV$ is expected to be sufficiently small to provide a reasonable estimate of the behavior. 
\begin{figure}
	\includegraphics{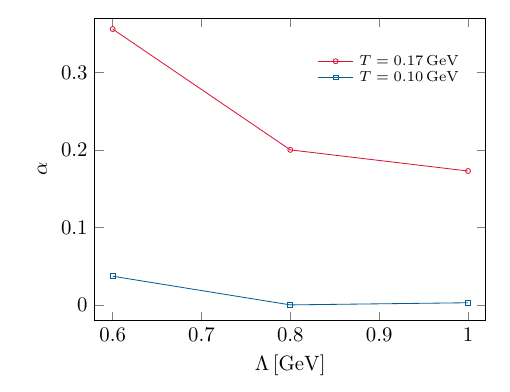}
	\caption{\label{fig:RG-consistency-cutoff-change} The \gls{RG} consistency measure~$\alpha$ as defined in~\cref{eq:RG-cons-cutoff}
	  evaluated for the minimum~$O=\bar{\sigma}_{\text{gs}}$ as a function of the \gls{UV} cutoff for the three-dimensional Litim regulator at $\mu=0$ and temperatures $T=0.1\GeV$ and $T=0.17\GeV$.}
\end{figure}
In \cref{fig:RG-consistency-cutoff-change}, we show the resulting values of $\alpha$. 
As already anticipated from the behavior of $\partial_{\bar{\sigma}} U$, we observe that $\alpha$ is smaller at lower temperatures.
Furthermore, we find that this quantity tends to decrease with increasing \gls{UV} cutoff.
We conclude that \gls{RG} consistency is rather well preserved for $T\lesssim0.1\GeV$, while \gls{UV}-cutoff artifacts start to play a significant role for $ T \gtrsim 0.1\GeV$ within the considered range of \gls{UV}-cutoffs.
\begin{figure*}
	\subfloat[\label{fig:sigma_vs_u_great_mu}%
	]{%
		\includegraphics{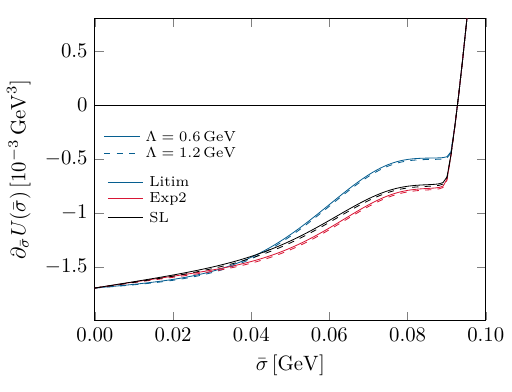}%
	}\hfill
	\subfloat[\label{fig:sigma_vs_u_greater_mu}%
	]{%
		\includegraphics{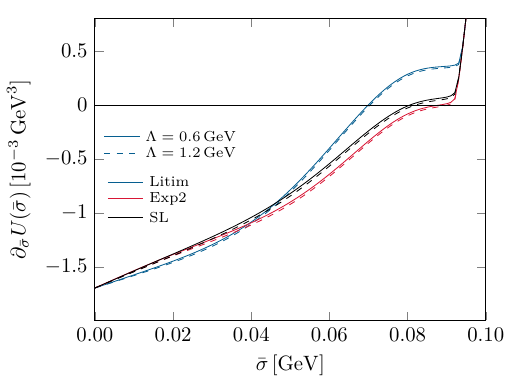}%
	}
	\caption{\label{fig:sigma_vs_u_high_mu} Derivative of the effective potential at $T=0.01\GeV$ and $\mu=0.3\GeV$ (left panel) and $\mu=0.312\GeV$ (right panel) for different regulator and \gls{UV}-cutoff combinations.}
\end{figure*}

Next, we consider large chemical potentials $\mu$ and small temperatures $T$ in \cref{fig:sigma_vs_u_high_mu}. 
More precisely, \cref{fig:sigma_vs_u_great_mu} displays $\partial_{\bar{\sigma}} U$ at $\mu=0.3\GeV$ and $T=0.01\GeV$, while \cref{fig:sigma_vs_u_greater_mu} corresponds to $\mu=0.312\GeV$ at the same temperature. 
In \cref{fig:sigma_vs_u_great_mu}, we observe that our results for $\partial_{\bar{\sigma}} U$ are in remarkable agreement around $\bar{\sigma}_\mathrm{gs}$ for the different \gls{UV}-cutoff and regulator combinations, despite the rather large chemical potential, which reflects the Silver-Blaze property. 
For $\bar{\sigma}<\bar{\sigma}_\mathrm{gs}$, however, a characteristic structure emerges in $\partial_{\bar{\sigma}} U$, indicating proximity to a first-order phase transition.
In this $\bar{\sigma}$-region, the results from the Litim regulator differ significantly from those obtained from the regulators Exp2 and SL. 
Contrary to the high-temperature case, the \gls{UV}-cutoff dependence is comparatively weak within the range of \gls{UV} cutoffs considered here.
A slight increase of the chemical potential to $\mu=0.312\GeV$, see \cref{fig:sigma_vs_u_greater_mu}, then causes a jump of $\bar{\sigma}_\mathrm{gs}$ due to the form of $\partial_{\bar{\sigma}} U$, signaling the onset of a first-order transition. 
Also, the results for $\partial_{\bar{\sigma}} U$ in the vicinity of $\bar{\sigma}_\mathrm{gs}$ as obtained from the different regulators now strongly differ, especially between the Litim and the two other regulators. 
These observations point to a potentially significant regulator dependence in this $(\mu,T)$ regime. 
We conclude that \gls{RG} consistency of $\partial_{\bar{\sigma}} U$ around $\bar{\sigma}_\mathrm{gs}$ is maintained up to the onset of the first-order phase transition. 
Beyond this transition regulator artifacts become nonnegligible.
We explicitly verified that the differences between results obtained with different regulators are larger than the changes induced by variations of $k_\mathrm{IR}$.
However, the $\mu$ value at which the regime associated with a very small slope of~$\partial_{\bar{\sigma}} U$ crosses zero is sensitive to small modifications of $\partial_{\bar{\sigma}} U$, such as those arising from variations of $k_\mathrm{IR}$.
From a phenomenological standpoint, this is relevant since this value of~$\mu$ determines the position of the first-order transition.

Finally, we turn to an analysis of \gls{RG} consistency of the $(\mu,T)$ phase diagram, employing the observable $\bar{\sigma}_\mathrm{gs}$ as order parameter. 
Due to the explicit chiral symmetry breaking, $\bar{\sigma}_\mathrm{gs}$ never vanishes exactly, and thus no sharp phase boundary exists in the strict sense.
Nevertheless, meaningful comparisons can be made between contour lines of constant $\bar{\sigma}_\mathrm{gs}$ for different regulators and \gls{UV} cutoffs.
In the following we employ $k_\mathrm{IR}=80\MeV$ in the computation of contour lines in order to keep the computational cost manageable.
We have verified that this choice is adequate for contour lines at low chemical potential $\mu$ and high temperature $T$, whereas at high $\mu$ and low $T$ the contour lines may exhibit slight modifications due to the effect discussed in the previous paragraph. 
However, we observe a consistent convergence behavior of $\partial_{\bar{\sigma}} U$ as  $k_\mathrm{IR}$ is lowered for the different regulators, and therefore expect these changes to also be consistent among the considered regulators. 
\begin{figure}
	\includegraphics{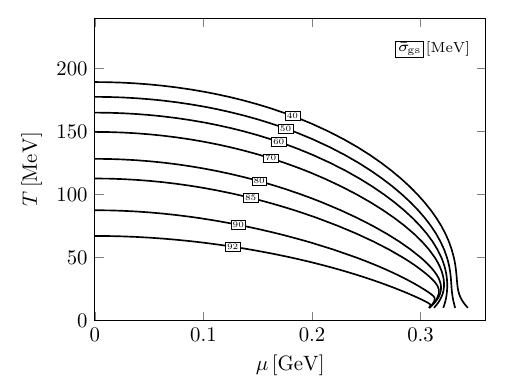}
	\caption{\label{fig:Litim-phase-dia} Phase diagram of the \gls{QMM}. Shown are contour lines for the order parameter $\bar{\sigma}_\mathrm{gs}$, exemplarily for the Litim regulator and $\Lambda=1\GeV$.}
\end{figure}
As an example, we show the contour lines obtained with the Litim regulator shape function at $\Lambda=1\GeV$ in \cref{fig:Litim-phase-dia}.
We restrict ourselves to $T>0.01\GeV$, since computations at lower temperatures become increasingly challenging from a numerical standpoint.
At small $\mu$, the contour lines reflect the expected crossover behavior: $\bar{\sigma}_\mathrm{gs}$ decreases continuously with increasing $T$ in agreement with our previous findings for $\partial_{\bar{\sigma}} U$ at small $\mu$ and high $T$. 
At small $T$, we find that $\bar{\sigma}_\mathrm{gs}$ remains nearly constant with increasing $\mu$ up to $\mu \approx 0.3\GeV$~\cite{Marko:2014hea,Khan:2015puu,Braun:2020bhy}, reflecting the Silver-Blaze property~\cite{Cohen:2003kd}.
Beyond this threshold, contour lines merge, indicating a rapid drop of $\bar{\sigma}_\mathrm{gs}$, which is characteristic of a first-order phase transition. 
This observation is in agreement with our results for $\partial_{\bar{\sigma}} U$ at high $\mu$ and small $T$.
Furthermore, we observe that the first-order line bends towards smaller $\mu$ as $T$ decreases, a phenomenon commonly referred to as back-bending~\cite{Schaefer:2004en}, see also Refs.~\cite{Tripolt:2017zgc,Zhang:2017icm,Aoki:2015mqa}.  
The Clausius-Clapeyron relation relates this fact to the appearance of negative entropy densities in this region of the phase diagram~\cite{Tripolt:2017zgc}.
Note that due to the explicit chiral symmetry breaking the back-bending region is shifted towards larger $\mu$ values compared to computations with identical \gls{UV} parameters and $c=0$. 

To assess \gls{RG} consistency of the $(\mu,T)$ phase diagram, we compare selected contour lines for various regulators and \gls{UV} cutoffs.
\begin{figure}
	\includegraphics{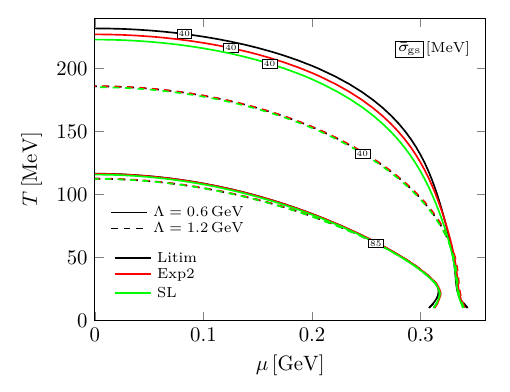}
	\caption{\label{fig:comparison-phase-dia} Comparison of selected contour lines ($\bar{\sigma}_\mathrm{gs}=40\MeV$ and $\bar{\sigma}_\mathrm{gs}=85\MeV$) corresponding to different combinations of regulator and \gls{UV} cutoff.}
\end{figure}
In \cref{fig:comparison-phase-dia} we show the $\bar{\sigma}_\mathrm{gs}=0.04\GeV$ and $\bar{\sigma}_\mathrm{gs}=0.085\GeV$ contours for all three regulators (Litim, Exp2, SL) and the considered minimal and maximal values of the \gls{UV} cutoff ($\Lambda=0.6\GeV$ and $\Lambda=1.2\GeV$).
At small $\mu$ and large $T$, \gls{UV}-cutoff artifacts dominate, and both \gls{UV}-cutoff and regulator dependences increase with increasing temperature, consistent with our analysis of $\partial_{\bar{\sigma}} U$ above. 
For $\bar{\sigma}_\mathrm{gs}=0.04\GeV$, deviations under a change of the \gls{UV} cutoff are significant.
A larger \gls{UV} cutoff generally shifts contour lines downward to smaller $T$.
At large $\mu$ and small $T$, in the region of the back-bending phenomenon, we instead find only a mild \gls{UV}-cutoff but a stronger regulator dependence. 
Although regulator effects appear to be strong in \cref{fig:sigma_vs_u_greater_mu}, the apparent fast variation of $\partial_{\bar{\sigma}} U$ with $\mu$ implies that the resulting contour lines differ only moderately. 
Still, the deviation of the Litim-regulator contour lines from Exp2 and SL is more pronounced than the mutual differences between Exp2 and SL. 
Moreover, the latter two regulators reduce the bending of the $\bar{\sigma}_\mathrm{gs}=0.085\GeV$ contour line towards small $\mu$, suggesting that at least part of the back-bending phenomenon may be a regulator artifact.
This interpretation is consistent with the findings in Ref.~\cite{Otto:2022jzl}, where mass-like regulators were employed.
Nevertheless, since the back-bending phenomenon is observed for all three regulators and all considered \gls{UV} cutoffs, we expect it to be a robust feature of the \gls{QMM} in the \gls{LPA} with three-dimensional regulators, irrespective of their smoothness and their classification by the principle of strongest singularity.

\section{Conclusions}\label{sec:conc}
We have discussed the predictive power of the \gls{QMM}, based on an analysis of its dependence on the \gls{UV} cutoff and regulator. 
To this end, we have defined a nontrivial parameter-fixing procedure that allows for a meaningful comparison of results obtained from different regulators and \gls{UV} cutoffs, i.e., to assess \gls{RG} consistency. 
In our analysis, we have considered a range of \gls{UV}-cutoff values, including those typically used in phenomenological studies of low-energy \gls{QCD}. 
The considered regulators come with different analytic properties and were chosen such that they cover a wide range in the space of regulator functions according to the principle of strongest singularity~\cite{Zorbach:2024zjx}. 
Based on this setup, we analyzed \gls{RG} consistency of the \gls{QMM}, with an emphasis on the phase diagram in the plane spanned by the temperature and quark chemical potential. 
For temperatures $T \lesssim 0.1\GeV$ and quark chemical potentials $\mu \lesssim 0.3\GeV$, the dependence of observables on the \gls{UV} cutoff and the regulator turns out to be weak. 
This does not come unexpected since the temperatures and chemical potentials are small compared to the considered \gls{UV} cutoffs in this regime. 
Therefore, dependences on the \gls{UV} cutoff and the regulator should be suppressed, provided the model parameters have been fixed consistently. 

At high temperatures and low chemical potentials, we have observed that the dependence of observables on the \gls{UV} cutoff is significantly stronger than the dependence on the regulator, at least within the considered range of \gls{UV} cutoffs. 
At low temperatures and high chemical potentials, beyond the Silver-Blaze threshold, the situation is different. 
In this regime, changes in physical observables as induced by a variation of the regulator are stronger than those induced by a variation of the \gls{UV} cutoff in the considered range. 
This regulator dependence leaves its imprint in the chiral phase structure. 
In particular, it quantitatively affects the back-bending phenomenon of the chiral first-order phase transition. 
Qualitatively, however, this phenomenon persists at least for all regulators and \gls{UV}-cutoff scales considered in this work, provided the parameter fixing is done consistently. 

To gain an even deeper understanding of the predictive power of the \gls{QMM}, it may be interesting to determine the \gls{UV} behavior in more detail. 
For example, it may be possible to choose even significantly larger values of the \gls{UV} cutoff than typically chosen based on phenomenological arguments. 
Note that the \gls{UV} cutoff is only an auxiliary quantity introduced to render loop integrals finite. 
While the \gls{UV} cutoff can be removed completely in, e.g., a one-loop calculation of the \gls{QMM} in the large-$N_{\rm c}$-limit~\cite{Braun:2018svj}, and possibly even in certain approximations beyond that~\cite{forthcomingpaper}, it may not be possible for the full theory, at least from the standpoint of scalar field theories in four dimensions, which are not expected to have a nontrivial continuum limit. 
In any case, it may be beneficial to consider optimization of the class of three-dimensional regulators typically employed in such model studies by, e.g., applying the principle of strongest singularity. 
Applying a consistent parameter-fixing procedure, it would also be interesting to go beyond \gls{LPA} since our results indicate that the back-bending phenomenon implying a negative entropy density in \gls{LPA} is neither a regulator nor a \gls{UV}-cutoff artifact.
This may be resolved at higher orders in the derivative expansion. 

\acknowledgements
We thank A.~Geissel, A.~K{\"o}nigstein, J.~M.~Pawlowski, F.~Rennecke, and S.~T{\"o}pfel for useful discussions and comments on the manuscript. 
D.H.R.~thanks the Department of Modern Physics of the University of Science and Technology of China (USTC) at Hefei for its hospitality, where part of this work was done.
This work has been supported in part by the \textit{Deutsche Forschungsgemeinschaft} (DFG, German Research Foundation) project-ID 279384907 – SFB 1245 (J.B., J.S.), 
the \textit{Deutsche Forschungsgemeinschaft} (DFG, German Research Foundation) through the CRC-TR 211 ``Strong-interaction matter under extreme conditions'' -- project number 315477589 -- TRR 211 (J.B., D.R., N.Z.,L.K.), and by the State of Hesse within the Research Cluster ELEMENTS (Project No.~500/10.006).

\appendix

\vfill
\section{Numerical treatment of the flow
equations} \label{app:num}
We implemented the Kurganov-Tadmor scheme in a semi-discrete way, i.e., we discretized the field space, but used a continuous \gls{RG} scale, which leads to a coupled system of ordinary differential equations. 
For all plots we used an extent in field space of $\bar{\sigma}_\mathrm{max}/\Lambda=1.5$ and a resolution of $\Delta \bar{\sigma} = 0.001\GeV$.
With respect to software, we employed Python's \textit{solve\_ivp}~\cite{2020SciPy-NMeth} in \textit{Python 3}~\cite{10.5555/1593511} (using various libraries~\cite{2020SciPy-NMeth,Hunter:2007,harris2020array}) to solve this system with ``LSODA'' as numerical time stepper with the parameters $rtol=atol=10^{-12}$ for its relative and absolute error.
For a detailed description of the numerical implementation of the Kurganov-Tadmor scheme within the \gls{fRG} framework, we refer the reader to, e.g., Refs.~\cite{Stoll:2021ori,Zorbach:2024rre,Koenigstein:2021syz,Koenigstein:2021rxj,Steil:2021cbu}.
The momentum integrals are solved using Numpy's \textit{polynomial.laguerre.laggauss} with $deg=100$.

\bibliography{bib}

\end{document}